\documentclass{svproc}
\usepackage{url}

\usepackage[pdftex]{graphicx}
\usepackage{amsmath}
\usepackage{amssymb}

\begin{document}
\mainmatter              
\title{Dipole-dipole interactions between neutrons}
\titlerunning{Dipole-dipole interactions between neutrons}  
%
\author{Renato Higa\inst{1} \and James F. Babb\inst{2} \and 
Mahir S. Hussein\inst{3,1,4}}
\authorrunning{R. Higa et al.} 
%
\tocauthor{Renato Higa, James F. Babb and Mahir S. Hussein}
\institute{Instituto de F\'\i sica, Universidade de S\~ao Paulo, R. do Mat\~ao 
1371, 05508-090,\\ S\~ao Paulo, SP, Brazil,
\and
ITAMP, Center for Astrophysics \textbar \ Harvard \& Smithsonian, 
MS 14, 60 Garden St., Cambridge, MA 02138, USA, 
\and
Instituto de Estudos Avan\c{c}ados, Universidade de S\~{a}o Paulo, 
C. P. 72012, 05508-970 S\~{a}o Paulo-SP, Brazil,
\and
Departamento de F\'{i}sica, Instituto Tecnol\'{o}gico de Aeron\'{a}utica, 
CTA,\\ S\~{a}o Jos\'{e} dos Campos, S.P., Brazil
}

\maketitle              

\begin{abstract}
In this work we present results of the dipole-dipole interactions between 
two neutrons, a neutron and a conducting wall, and a neutron between two 
walls. As input, we use dynamical electromagnetic dipole polarizabilities 
fitted to chiral EFT results up to the pion production threshold and at the 
onset of the Delta resonance. 
Our work can be relevant to the physics of confined ultracold neutrons 
inside bottles. 
\end{abstract}

\section{Introduction}

The Casimir effect is a quite popular example of a non-trivial phenomenon 
arising from quantum fluctuations on the vacuum energy. The Casimir-Polder 
force was originaly devised to address the mismatch of the van der Waals 
$1/r^6$ tail of interatomic interactions and observations of Overbeek on 
quartz powder in colloid suspension. The correct asymptotic $1/r^7$ behavior 
is obtained by taking into account 
retardation effects due to the finiteness of the speed of light. 
Shortly after, Casimir related this force to quantum fluctuations 
of the vacuum between two neutral objects. Such force should appear due 
to the change in the zero-point electromagnetic energy between two neutral, 
conducting plates, an experimentally confirmed fact since then 
(see~\cite{milonni-shih-92} and references therein). 

A huge body of work has been devoted to this subject in atomic 
physics~\cite{Bab10}. Feinberg and Sucher~\cite{FeiSuc70} derived the 
Casimir-Polder (CP) interaction between two neutral spinless particles 
via the two-photon exchange mechanism and general analytical properties 
of the related Compton scattering sub-amplitudes. The result is given in 
terms of atomic dipole polarizabilities reflecting the linear response 
to an external electromagnetic field. 
Arnold~\cite{Arn73} was the first to calculate the CP interaction between 
two neutrons, however, at that time only the static, electric dipole 
polarizability data were available with nowadays outdated values. 
We extended Arnold's idea~\cite{BabbHigaHussein17,HBH17-2} to include 
dynamic electric and magnetic dipole polarizabilities with updated 
information from low-energy chiral effective field theory analysis. 
We also performed calculations of the CP-interaction between a neutron 
and a wall, and one neutron between two walls. 
In the following we 
summarize our main results and present an outlook for future studies. 

\section{Compton scattering and neutron polarizabilities}

\begin{figure}
\begin{tabular}{lr}
\includegraphics[width=0.5\textwidth]{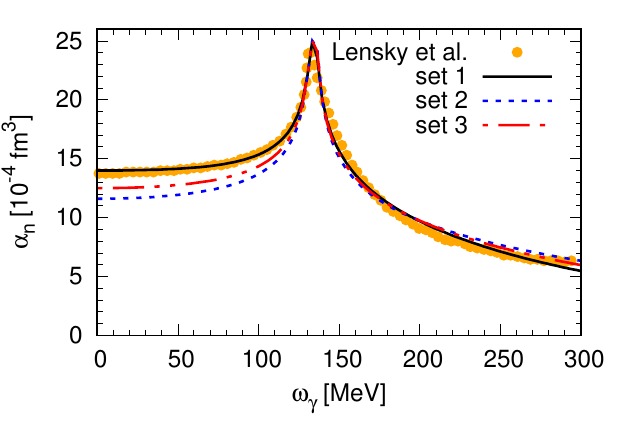}&
\includegraphics[width=0.5\textwidth]{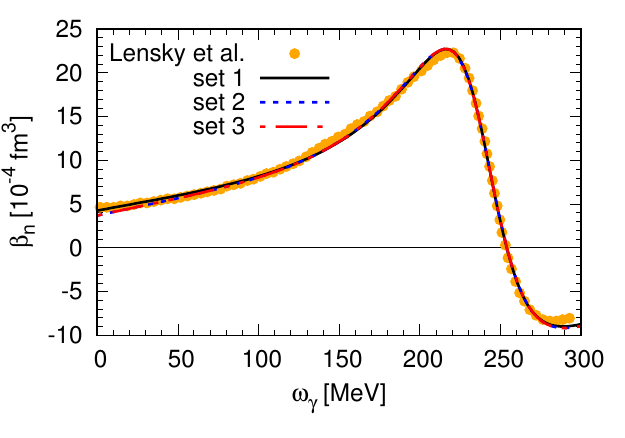}
\end{tabular}
\caption{
Dynamic electric (left) and magnetic (right) polarizabilities, as functions
of the photon energy $\omega_{\gamma}$.
The yellow circles are $\chi$EFT results of Lensky
{\em et al.}~\cite{Lensky:2015awa} while sets 1, 2, and 3 correspond to
our parametrizations using the numbers specified in Table~\ref{tab:pol-par}.
\label{fig:n-polariz}
}
\end{figure}

Compton scattering on both proton and neutron became a wide subject in 
hadron physics comprising many theoretical and experimental efforts around 
the world. See recent review~\cite{Hagelstein:2015egb} for the current status 
of this line of research, where one also finds the intricate details on 
how to extract information on the polarizabilities from the Compton 
scattering amplitudes. Chiral effective field theory ($\chi$EFT), the 
effective theory of the underlying strong interactions (QCD), has being 
used to make rigorous and systematic predictions to Compton scattering 
observables at photon energies around and below the $\Delta$-resonance 
excitation energy. The most recent $\chi$EFT calculations of Lensky 
{\em et al.}~\cite{Lensky:2015awa} 
for the electric $\alpha_n(\omega)$ and magnetic $\beta_n(\omega)$ dynamical 
dipole polarizabilities of the neutron 
is shown in Fig.~\ref{fig:n-polariz} 
together with our low-energy parametrizations 
\begin{eqnarray}
&&\alpha_n(\omega)=\frac{\alpha_n(0)\sqrt{(m_{\pi}\!+\!a_1)
(2M_n\!+\!a_2)}(0.2a_2)^2}{\sqrt{(\sqrt{m_{\pi}^2\!-\!\omega^2}\!+\!a_1)
(\sqrt{4M_n^2\!-\!\omega^2}\!+\!a_2)}\big[|\omega|^2\!+\!(0.2a_2)^2\big]},
\label{eq:edip-polariz1}
\\[0.0cm]
&&\beta_n(\omega)=\frac{\beta_n(0)\!-\!b_1^2\omega^2\!+\!b_2^3\,
{\rm Re}(\omega)}{(\omega^2\!-\!\omega_{\Delta}^2)^2\!+\!
|\omega^2\Gamma_{\Delta}^2|}, 
\label{eq:mdip-polariz1}
\end{eqnarray}
with the set of parameters from Table~\ref{tab:pol-par} 
(more details in~\cite{BabbHigaHussein17}). 

\begin{table}[tbh]
\caption{Parameters of Eqs.~(\ref{eq:edip-polariz1}), (\ref{eq:mdip-polariz1})
fitted to the theoretical curves of Ref.~\cite{Lensky:2015awa}.
$\alpha_n(0)$ and $\beta_n(0)$ units are $10^{-4}{\rm fm}^3$, the remaining 
ones in MeV.
\label{tab:pol-par}}
\begin{center}\begin{tabular}{c||c|c|c|c|c|c|c|c}
& $\alpha_n(0)$ & $a_1$ & $a_2$ &$\beta_n(0)$ & $b_1$ & $b_2$ &
$\omega_{\Delta}$ & $\Gamma_{\Delta}$ \\ \hline
Set 1 & 13.9968 & 12.2648 & 1621.63 & 4.2612 & 8.33572 & 22.85 & 241.484 & 66.92
65 \\ \hline
Set 2 & 11.6 & 2.2707 & 2721.47 & 3.7 & 8.67962 & 24.2003 & 241.593 & 68.3009 \\
 \hline
Set 3 & 12.5 & 5.91153 & 2118.79 & 2.7 & 9.27719 & 26.328 & 241.821 & 70.8674 \\
\end{tabular}\end{center}
\end{table}

\section{Casimir-Polder interactions with neutrons}

The CP interaction between two neutrons is given 
by~\cite{FeiSuc70,Bab10,SprKel78,BabbHigaHussein17} 
\begin{eqnarray}
&&V_{CP,nn}(r)=-\frac{\alpha_0}{\pi r^6 }\,I_{nn}(r)\,,
\nonumber\\[0mm]
&&I_{nn}(r)=\int_0^{\infty}d\omega\,e^{-2\alpha_0\omega r}
\Big\{\Big[\alpha_n(i\omega)^2+\beta_n(i\omega)^2\Big]P_E(\alpha_0\omega r)
\nonumber\\[0mm]
&&\hspace{3.0cm}
+\Big[\alpha_n(i\omega)\beta_n(i\omega)
+\beta_n(i\omega)\alpha_n(i\omega)\Big]
P_M(\alpha_0\omega r)\Big\}\,,
\nonumber\\[0mm]
&&P_E(x)=x^4+2x^3+5x^2+6x+3\,,\quad P_M(x)=-(x^4+2x^3+x^2)\,,
\end{eqnarray}
where $\alpha_0\approx 1/137$ is the electromagnetic fine structure constant. 

For the neutron-Wall (nW) CP potential one 
has~\cite{ZhouSpruch95,khar97,BabbHigaHussein17}
\begin{eqnarray}
&&V_{CP,nW}(r) = -\frac{\alpha_0}{4\pi r^3}J_{nW}(r)\,,
\quad J_{nW}(r)=\int_0^{\infty}d\omega\,e^{-2\alpha_{0}\omega r}
\alpha_{n}(i\omega)Q(\alpha_{0}\omega r)\,,
\nonumber\\[0mm]
&&Q(x)=2x^2+2x+1\,.
\label{eq:integ_nW}
\end{eqnarray}

Finally, for two Walls separated by a distance $L$ and one neutron 
in between, at a distance $z$ from the 
midpoint~\cite{ZhouSpruch95,khar97,BabbHigaHussein17}, 
\begin{align}
&V_{CP,WnW}(z, L)=
\nonumber\\
&-\frac{1}{\alpha_{0}\pi L^4}\int_{0}^{\infty}u^3du\,
\alpha\left(i\frac{u}{\alpha_{0}L}\right)
\int_{1}^{\infty}\frac{dv}{\sinh(uv)}\left[
v^2\cosh\left(\frac{2z}{L}uv\right)-e^{-uv}\right]\,.
\label{eq:Vwnw01}
\end{align}

\section{Results}

\begin{figure}[tbh]
\begin{tabular}{lr}
\includegraphics[width=0.57\textwidth,clip]{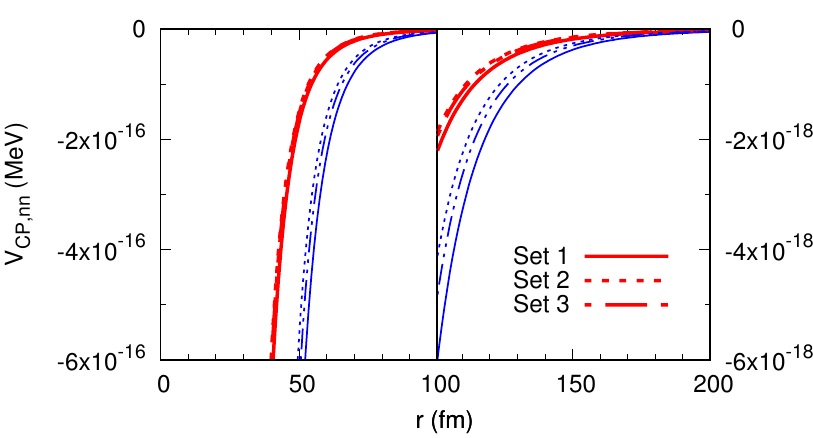}&
\includegraphics[width=0.43\textwidth,clip]{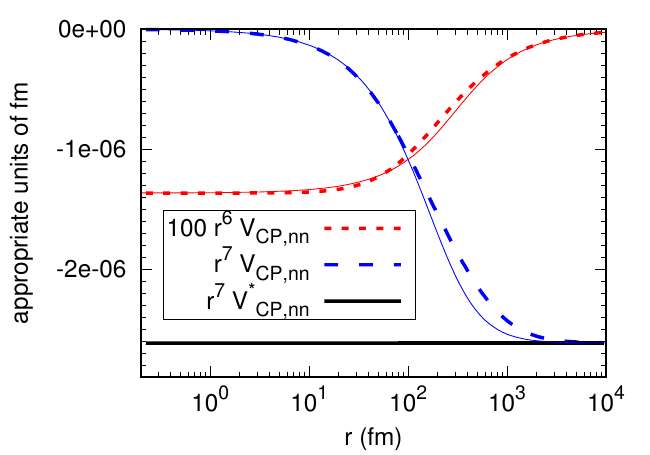}\\
\includegraphics[width=0.57\textwidth,clip]{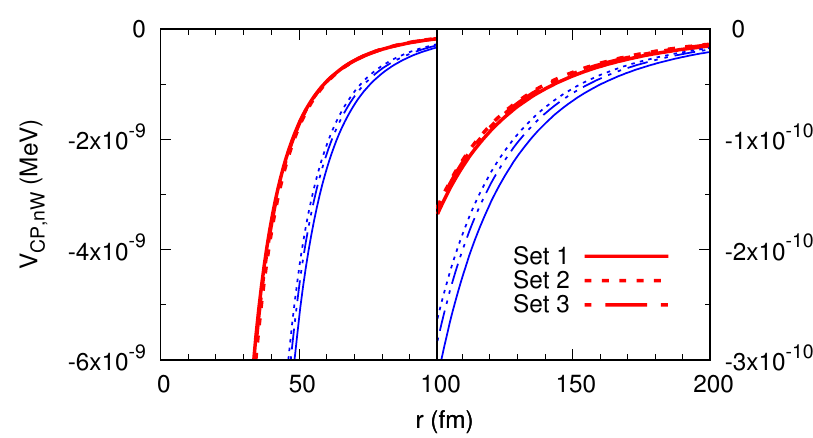}&
\includegraphics[width=0.43\textwidth,clip]{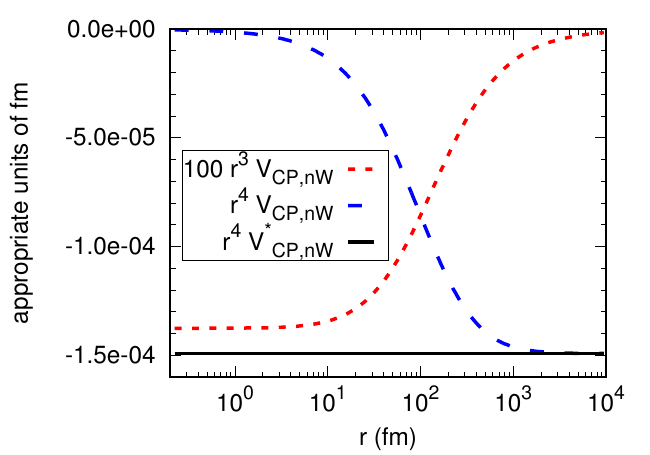}\\
\includegraphics[width=0.49\textwidth,clip]{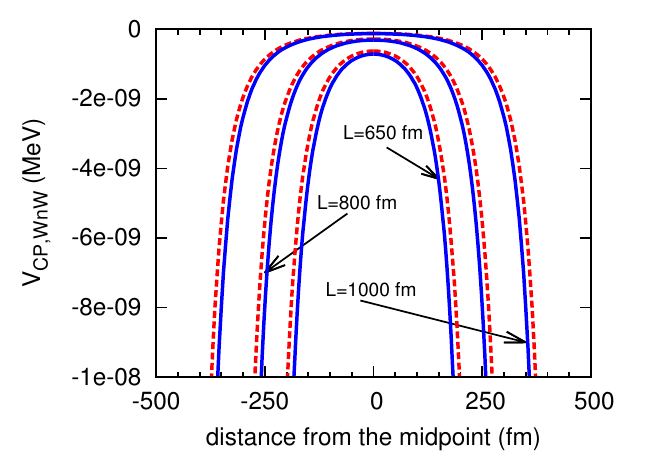}&
\includegraphics[width=0.49\textwidth,clip]{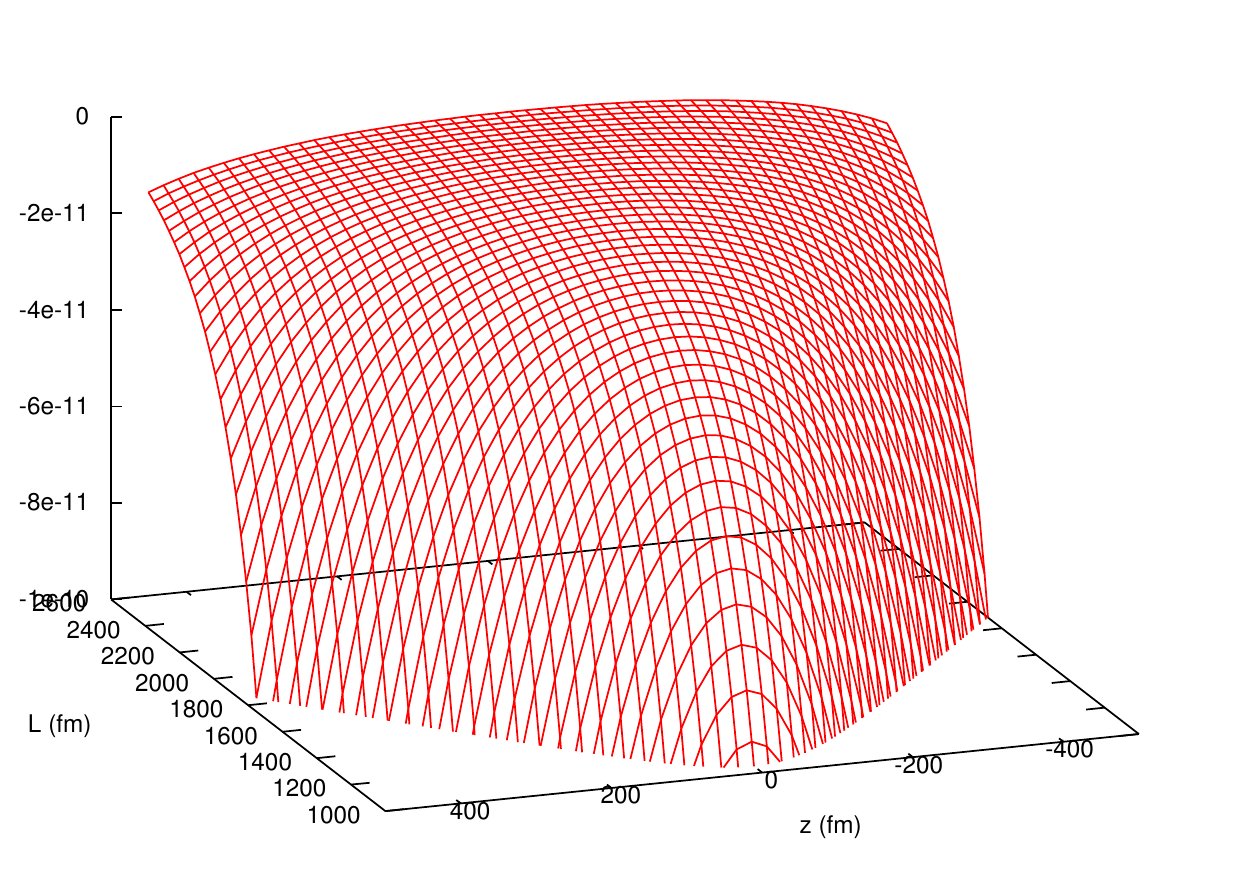}
\end{tabular}
\caption{\protect
Results for the various CP-interactions. See text for details.}
\label{fig:Vcp2}
\end{figure}

Our results are shown in Fig.~\ref{fig:Vcp2}. 
First row, for $V_{CP,nn}$ as function of the separation distance $r$. 
In the left panel, the (red) curves with smaller magnitudes stand for 
dynamical, $\omega$-dependent polarizabilities while the (blue) ones with 
higher magnitudes stand for the static limit. On the right panel, the 
(red/blue) short-dashed/long-dashed curves are $V_{CP,nn}$ multiplied by 
appropriate factors ($100r^6$/$r^7$), the (red/blue) solid lines are the 
arctan parametrization~\cite{OCaSuc69} that phenomenologically makes the 
transition from short-distance
van der Waals to the asymptotic $1/r^7$ Casimir-Polder behavior~\cite{Arn73}. 
Second row, for $V_{CP,nW}$ and shows similar qualitative behavior as 
$V_{CP,nn}$. 
Third row, for $V_{CP,WnW}$ as function of $L$ and $z$. 
For ultracold neutrons, these attractive CP forces may compete with 
repulsive Fermi pseudopotential, {e.g.}, $V_F\approx 252\,{\rm neV}$ 
for Ni~\cite{BabbHigaHussein17}. 
%

%
%

\end{document}